\documentclass[11pt,twoside]{article}

\usepackage{galev06}
\usepackage{epsf}
\usepackage{psfig}
\usepackage{lscape}
\usepackage{comment}
\usepackage{wrapfig}

\begin{document}
\title{Relating Diffusion Properties of Cosmic-Ray Electrons to Star
  Formation Activity within Normal Galaxies}   %%% Fill in title 
\author{E.J. Murphy$^{1}$, G. Helou$^{2}$, L. Armus$^{2}$,
  R. Braun$^{3}$, J.D.P. Kenney$^{1}$, and the SINGS team}
\affil{$^{1}$Yale University; $^{2}${\it Spitzer} Science
  Center/Caltech; $^{3}$ASTRON}

%\title{}   %%% Fill in title
%\author{}   %%% Fill in author names
%\affil{}    %%% Fill in author affiliations

\begin{abstract} %%% Abstract to run on from here.
%%GHX: COMPRESS to letter-like phrasing!
Using data obtained as part of the {\it Spitzer} Infrared Nearby
Galaxies Survey \citep[SINGS;][]{rk03} and WSRT-SINGS radio continuum
survey \citep{rb06}, we study the effects of star-formation activity
on the far-infrared (FIR)--radio correlation {\it within} galaxies.
This is done by testing a phenomenological model for the correlation,
which describes the radio image as a smeared version of the FIR
image. 
We find that this description works particularly well for galaxies
with higher infrared surface brightnesses, yielding best-fit smoothing
scale-lengths of a few hundred parsecs, substantially shorter than
those for lower surface brightness galaxies.
We interpret this result to suggest that galaxies with higher disk
averaged star formation rates have had a recent episode of enhanced
star formation and are characterized by a higher fraction of young
cosmic-ray (CR) electrons compared to galaxies with lower star
formation activity. 
\end{abstract}

%%% MAIN BODY OF TEXT GOES HERE. CONSULT "INSTRUCTIONS FOR AUTHORS USING
%%% LATEX2E MARKUP", SECTIONS 2.3-2.6 FOR HELP WITH EQUATIONS, FIGURES,
%%% AND TABLES.

\vskip -0pt
\vskip -12pt
\section*{Introduction}
\vskip -12pt
The relativistic phase of the interstellar medium (ISM) is a
collision-less gas of diffusing cosmic-ray (CR) particles. 
While this phase is often overlooked in extragalactic studies of star
formation, largely due to the difficulties involved with making
direct observations of CR nuclei, it has an energy density comparable
to the gaseous phases proving it to be equally important in
%%GXH: REMOVE "other" IN SENTENCE.
studies of galaxy evolution.
The distances diffused by CRs within galaxies depends on their age
%%GXH: SHOULD BE "The distances diffused by CRe..." INSTEAD OF
%%"diffusion scale-length" 
and effective mean free path. 
%%GXH: SHOULD BE "effective mean free path" INSTEAD OF "ability to diffuse"
Consequently, pinning down these scale-lengths can provide rough
estimates for the ages of related star formation episodes as well as
help to constrain associated ISM parameters.
%associated with the propagation characteristics of CR electrons.

Synchrotron emission from normal galaxies is readily observable
%%GXH:  REMOVE "often"
at radio wavelengths and directly probes the coupling between a
galaxy's distribution of CR electrons and its magnetic field.
However, these data alone do not provide information on the
propagation history of CR electrons.
The close spatial correlation between the far-infrared (FIR) and
predominantly non-thermal radio continuum (RC) emission of normal
late-type galaxies suggests a common physical origin, most likely
through the processes of massive star formation
\citep[e.g.][and references therein]{ejm06a}.
Massive stars
%($M\geq8~M_{\sun}$) 
are thought to be the primary heating sources of dust and whose
remnants accelerate CR electrons in galaxies. 
Since CR electrons diffuse measurably farther ($\sim$1~kpc) than the
mean free path of dust-heating photons ($\sim$100~pc), 
%%GXH: THIS SHOULD BE <=100pc, NOR comparable to, and 1kpc IS ALSO TOO LARGE
a galaxy's radio image should appear as a smoother version of its FIR
map which can be related by the diffusion history of its CR electrons
\citep{bh90}.
%\citet{bh90} suggested that a galaxy's RC image should appear as
%smoother version of its FIR map and be related by the diffusion
%history of its CR electrons. 

\begin{wrapfigure}{R}{6.5cm}
\vskip -14pt
  \resizebox{6.5cm}{!}{
    \plotone{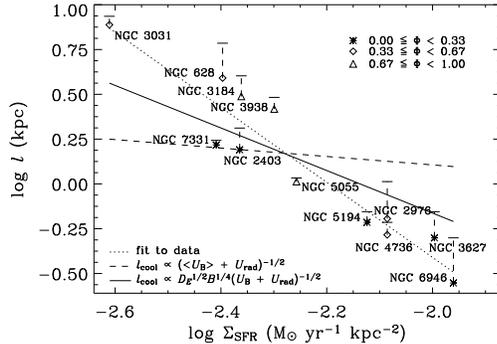}}
\vskip -10pt
\hskip -18pt
\parbox{0.6\textwidth}{\caption{\scriptsize
    The SFR surface densities, ($\Sigma_{\rm SFR}$) 
    are plotted against the best-fit smoothing kernel scale-lengths.
    The improvement in the spatial correlation between the
    70~$\micron$ and 22~cm maps ($\Phi$) is indicated by the
    plotting symbol.  
    Also plotted is the fit to the data ({\it dotted line}) and the
    expected diffusion scale-lengths due to inverse Compton (IC)
    losses in a fixed magnetic field ({\it dashed line}) and
    synchrotron + IC losses with an energy-dependent diffusion
    coefficient for the steepest index ({\it solid line}).   
    The increase to the best-fit scale-length using 22~cm maps
    corrected for free-free emission is indicated by a vertical
    line.} 
  \label{SFRDleg}}
%\vskip -75pt
\vskip -35pt
\end{wrapfigure}

Using this phenomenology, we relate the spatial distributions of FIR
(70~$\mu$m) and RC (22~cm) emission for 12 spiral galaxies by a
smearing kernel which, when convolved with a galaxy's 70~$\mu$m map,
minimizes the residuals between the spatial distributions of the two
maps. 
In Figure \ref{SFRDleg} we plot the best-fit smearing scale-length
against each galaxy's total-infrared (TIR; 3-1100~$\mu$m) surface
brightness, which we express as star formation rate (SFR) surface
density ($\Sigma_{\rm SFR}$).
We find the general property that galaxies are better modeled by
larger smearing scale-lengths if their disk-averaged SFR is lower.
The complete study can be found in \citet{ejm06b}.

\vskip -6pt
\vskip -6pt
\section*{Discussion}
\vskip -12pt
%The physical interpretation of our results indicates that CR electrons
We interpret our results as an indication that CR electrons
%%GXH: CHANGE TO "We interpret our results as an indication that CR..."
are, on average, closer to their place of origin in galaxies having
higher star formation activity. 
%%GXH: NEXT SENTENCE, SIMPLIFY TO "We consider three general explanations
%%for this result, whereby CR electrons in more active galaxies 
%%(1) have shorter lifetimes due to a high energy loss rate;
%%(2) AS IS...; or
%%(3) have been accelerated more recently and are younger on average.
We consider three general explanations for this result, whereby CR
electrons in more active galaxies:  
(1) have shorter lifetimes due to a high energy loss rate;
(2) diffuse more slowly due to a high ISM density and magnetic
field strength; or
(3) have been accelerated more recently and are younger on average.
%Since the diffusion scale-length of CR electrons depends on their age
%and ability to diffuse through a galaxy's ISM, 3 general explanations
%are possible for this result:  CR electrons   
%(1) have relatively short lifetimes due to a high energy loss rate;
%(2) diffuse more slowly due to a high ISM density and magnetic
%field strength; or  
%(3) have been recently accelerated and are relatively young.
The first two of these explanations are compatible with
%%GXH: CHANGE "applicable in the case of " to "are compatible with..."
steady-state star formation, while the third 
suggests a recent episode of enhances star formation.

%%GXH: REWRITE THIS LAST PARA:
In Figure \ref{SFRDleg} we plot the expected relation between CR
electron diffusion length and TIR surface brightness for 
the steady-state cases, taking into account the dependence
of diffusion on the ISM parameters, and including Inverse Compton losses
(dashed line) and then both Inverse Compton and synchrotron losses
(solid line).
For the latter, we assume that CR electrons diffuse according to an
energy dependent diffusion coefficient which scales with ISM density
and magnetic field strength, and plot the steepest scaling
\citep[see][for details]{ejm06b}.  The observed trend is 
is $\sim$9 and $\sim$2 times steeper than the simple models shown in
Figure \ref{SFRDleg}.
We therefore conclude that galaxies with higher $\Sigma_{\rm SFR}$ are
likely to have experienced a recent episode of enhanced star formation
compared to 
%characterized by SFHs that have peaked more recently than
galaxies with lower disk-averaged SFRs, leading to a larger fraction
of relatively young CR electrons within their disks.

\acknowledgements %%% Text of acknowledgements runs on after this command.
%E.J.M. acknowledges support from the {\it Spitzer} Science Center.
As part of the {\it Spitzer} Legacy Science Program,
support was provided by NASA through Contract Number 1224769 issued
by JPL, Caltech, under NASA contract 1407.
%E.J.M. would like to acknowledge support for this work provided by
%the {\it Spitzer} Science Center Visiting Graduate Student program.
%As part of the {\it Spitzer} Space Telescope Legacy Science Program,
%support was provided by NASA through Contract Number 1224769 issued
%by the Jet Propulsion Laboratory, California Institute of Technology
%under NASA contract 1407.
\vskip -16pt
\vskip -12pt

%%% THE BIBLIOGRAPHY
%%%
%%% CONSULT SECTION 3 OF "INSTRUCTIONS FOR AUTHORS" FOR HOW TO USE NATBIB.
%%% AUTHORS ARE ENCOURAGED TO USE EITHER THE "THEBIBLIOGRAPY" ENVIRONMENT
%%% BY UNCOMMENTING (DELETING THE "%" SYMBOL) THE COMMANDS BELOW, OR BY
%%% USING THE BIBTEX ENVIRONMENT. TO FIND OUT WHICH IS APPLICABLE TO YOUR
%%% CONTRIBUTION, CONSULT THE VOLUME EDITORS FOR YOUR PROCEEDINGS.
%%%

\end{document}